\documentclass[aps,pra,twocolumn,showpacs,superscriptaddress,10pt]{revtex4-1}
\usepackage{amssymb,amsthm,amsmath,amsfonts}
\usepackage{color,graphicx,ulem}
\usepackage[colorlinks=true,urlcolor=blue,citecolor=blue,linkcolor=blue]{hyperref}

\begin{document}
 
\title{Is ``genuine multipartite entanglement'' really genuine?}

\author{S. Gerke}\email{stefan.gerke@uni-rostock.de}
\affiliation{Arbeitsgruppe Theoretische Quantenoptik, Institut f\"ur Physik, Universit\"at Rostock, D-18051 Rostock, Germany} 
\author{J. Sperling}
\affiliation{Arbeitsgruppe Theoretische Quantenoptik, Institut f\"ur Physik, Universit\"at Rostock, D-18051 Rostock, Germany} 
\author{W. Vogel}
\affiliation{Arbeitsgruppe Theoretische Quantenoptik, Institut f\"ur Physik, Universit\"at Rostock, D-18051 Rostock, Germany} 
\author{Y. Cai}
\affiliation{Laboratoire Kastler Brossel, UPMC-Sorbonne Universit\'es, CNRS, ENS-PSL Research University, Coll\`ege de France; 4 place Jussieu, 75252 Paris, France}
\author{J. Roslund}
\affiliation{Laboratoire Kastler Brossel, UPMC-Sorbonne Universit\'es, CNRS, ENS-PSL Research University, Coll\`ege de France; 4 place Jussieu, 75252 Paris, France}
\author{N. Treps}
\affiliation{Laboratoire Kastler Brossel, UPMC-Sorbonne Universit\'es, CNRS, ENS-PSL Research University, Coll\`ege de France; 4 place Jussieu, 75252 Paris, France}
\author{C. Fabre}
\affiliation{Laboratoire Kastler Brossel, UPMC-Sorbonne Universit\'es, CNRS, ENS-PSL Research University, Coll\`ege de France; 4 place Jussieu, 75252 Paris, France}
\date{\today}

\begin{abstract} 
	The existence of non-local quantum correlations is certainly the most important specific property of the quantum world. 
	However, it is a challenging task to distinguish correlations of classical origin from genuine quantum correlations, especially when the system involves more than two parties, for which different partitions must be simultaneously considered. 
	In the case of mixed states, intermediate levels of correlations must be introduced, coined by the name inseparability. 
	In this work, we revisit in more detail such a concept in the context of continuous-variable quantum optics. 
	We consider a six-partite quantum state that we have effectively generated by the parametric downconversion of a femtosecond frequency comb, the full $12 \times 12$ covariance matrix of which has been experimentally determined. 
	We show that, though this state does not exhibit "genuine entanglement'', it is undoubtedly multipartite-entangled. 
	The consideration of not only the entanglement of individual mode-decompositions but also of combinations of those solves the puzzle and exemplifies the importance of studying different categories of multipartite entanglement.
\end{abstract}
\maketitle

\section{Introduction}
	Entanglement appears nowadays as a major subject of research in quantum physics, long after the pioneering contributions of Einstein, Podolsky, Rosen~\cite{EPR35} and Schr\"odinger~\cite{SCH35}. 
	It is the main quantum resource in a vast number of applications in quantum information~\cite{BL05}.
	Entanglement witnesses can uncover such quantum correlations~\cite{HHHH09,GT09} in either discrete variables, when measurements are made using photon counters, or in continuous variables by employing homodyne detection, as far as quantum states of light are concerned.

	Pure entangled states have been first considered in bipartite systems.
	The case of mixed correlated states turns out to be more involved, and an intermediate situation between factorized and entangled states has been introduced, namely the separable states, which are statistical mixtures of factorized pure states~\cite{W89}.
	A number of entanglement probes for continuous-variable systems have been studied~\cite{HHHH09,GT09}, the partial transpose test being among the most popular ways to pinpoint inseparability. 
	These criteria are in most cases sufficient but not necessary to detect the different levels of correlation.
	The problem is simpler if one restricts oneself to bipartite Gaussian states, for which  the partial transposition of the covariance matrix is a necessary and sufficient entanglement identifier~\cite{P96,S00,DGCZ00}.

	The complexity of the separability problem increases by a large amount when one wants to tackle the case of three- or, more generally, multipartite systems.
	In these situations one has a rapidly increasing number of choices in the bunching of parties on which one searches for a possible factorization.
	Hence, the inseparability between the individual degrees of freedom exhibits a much richer and complex structure which begins to be studied~\cite{HV13,VLM14,SSV14}. 
	For example, the difference between bipartite and multipartite systems is highlighted by the existence of multimode Gaussian states whose entanglement cannot be uncovered by the partial transposition~\cite{WW01,DSHPES11}.

	As a special case, combinations of all possible bipartitions of the total system have been the subject of many studies.
	A state which is not a statistical mixture of bipartite factorized density matrices is called in the literature ``genuinely'' multipartite entangled.
	The detection of genuine entanglement is at the focus of attention~\cite{HHH01,WH05,YS05,HJ08,HHK09,HMGH10,SHYSRJ13,BBSNHMGB13,SL15}.
	This interest can be explained by the fact that genuine entanglement implies multipartite entanglement for every other separation of the modes.
	However, if a state does not exhibit this specific kind of entanglement, no further conclusions on other forms of multipartite quantum correlations can be drawn. 
	Thus, it is certainly indispensable to study what happens beyond genuine entanglement. 
	This is the subject of the present paper.

	On the experimental side of continuous-variable quantum optics, multipartite quantum correlated states have been first produced by mixing, in an appropriate way, different squeezed states on beam splitters~\cite{YULF15}.
	More recently, multimode Gaussian states (either spatial or frequency modes) have been directly generated by a multimode optical nonlinear device~\cite{PMSBO11,CMP14}.
	In the multi-frequency case, the experimental determination of the full covariance matrix of a ten-mode ``quantum frequency comb'' has allowed to uncover the complex structure of its quantum properties, and in particular the entanglement of all its possible partitions.~\cite{RMSFT14,ARCFFT14,GSVCRTF15}.

	In this paper, we will characterize states which are not genuinely entangled and yet exhibit a rich multipartite entanglement structure.
	In order to achieve this, we will formulate different notions of separability and entanglement, then we will provide a method to qualify them in a general case.
	Using such a theoretical method, we will uncover the structure of multimode entanglement in a six-mode Gaussian state that has been produced experimentally.
	Using multimode parametric downconversion of a femtosecond frequency comb source, we generate such a highly multimode quantum state of light that spans on many frequency modes.
	Even though this state does not exhibit genuine entanglement, it is shown to include all other forms of higher-order entanglement.

\section{Combinations of modal partitions}
	We consider multimode states which are based on a $N$-fold tensor product Hilbert space $\mathcal{H}=\mathcal{H}_1\otimes\ldots\otimes\mathcal{H}_N$, where $\mathcal{H}_j$ is the local Hilbert space of the $j$-th mode. 
	A particular $K$-partition $\mathcal{I}_1{:}\ldots{:}\mathcal{I}_K$ decomposes the set of modes, $\{1,\ldots,N\}$, into $K$ non-empty, disjoint subsets $\mathcal I_k$ (for $k=1,\ldots,K$).
	We will call such a partition an {\it individual} $K$-partition.

	The corresponding pure factorized states are product states, $|s_{\mathcal{I}_1{:}\ldots{:}\mathcal{I}_K}\rangle=|a_{\mathcal{I}_1}\rangle\otimes \ldots \otimes |a_{\mathcal{I}_K}\rangle$, consisting of states $|a_{\mathcal I_k}\rangle\in\bigotimes_{j\in\mathcal I_k}\mathcal H_j$.
	Subsequently, a continuous-variable, mixed $\mathcal{I}_1{:}\ldots{:}\mathcal{I}_K$-separable state is defined as
	\begin{equation}
		\hat\sigma_{\mathcal{I}_1{:}\ldots{:}\mathcal{I}_K}=\int dP(s_{\mathcal{I}_1{:}\ldots{:}\mathcal{I}_K})|s_{\mathcal{I}_1{:}\ldots{:}\mathcal{I}_K}\rangle\langle s_{\mathcal{I}_1{:}\ldots{:}\mathcal{I}_K}|,
		\label{Eq:IndPartSep}
	\end{equation}
	where $P$ is a classical probability distribution over the set of pure (continuous-variable) separable states.

	A state is called $K$-separable if it can be written as a statistical mixture of separable states with respect to the different $K$-partitions $\mathcal{I}_1{:}\ldots{:}\mathcal{I}_K$,
	\begin{equation}
		\hat\sigma_{K}=\sum_{\mathcal{I}_1{:}\ldots{:}\mathcal{I}_K} p_{\mathcal{I}_1{:}\ldots{:}\mathcal{I}_K}\hat\sigma_{\mathcal{I}_1{:}\ldots{:}\mathcal{I}_K},
		\label{Eq:KSeparable}
	\end{equation}
	where $p_{\mathcal{I}_1{:}\ldots{:}\mathcal{I}_K}$ are probabilities and $\hat\sigma_{\mathcal{I}_1{:}\ldots{:}\mathcal{I}_K}$ are the corresponding $\mathcal{I}_1{:}\ldots{:}\mathcal{I}_K$-separable states in Eq.~\eqref{Eq:IndPartSep}.
	We will refer to this combination of individual $K$-partitions as {\it convex combination} of $K$-partitions.
	A state is called $K$-entangled if it cannot be written in the manner specified in Eq.~\eqref{Eq:KSeparable}.
	In particular, a state which is not ``biseparable'' ($K=2$) is precisely the ``genuinely multipartite entangled state'' studied in the literature.

	\begin{figure}[ht]
		\centering
		\includegraphics[width=0.425\textwidth]{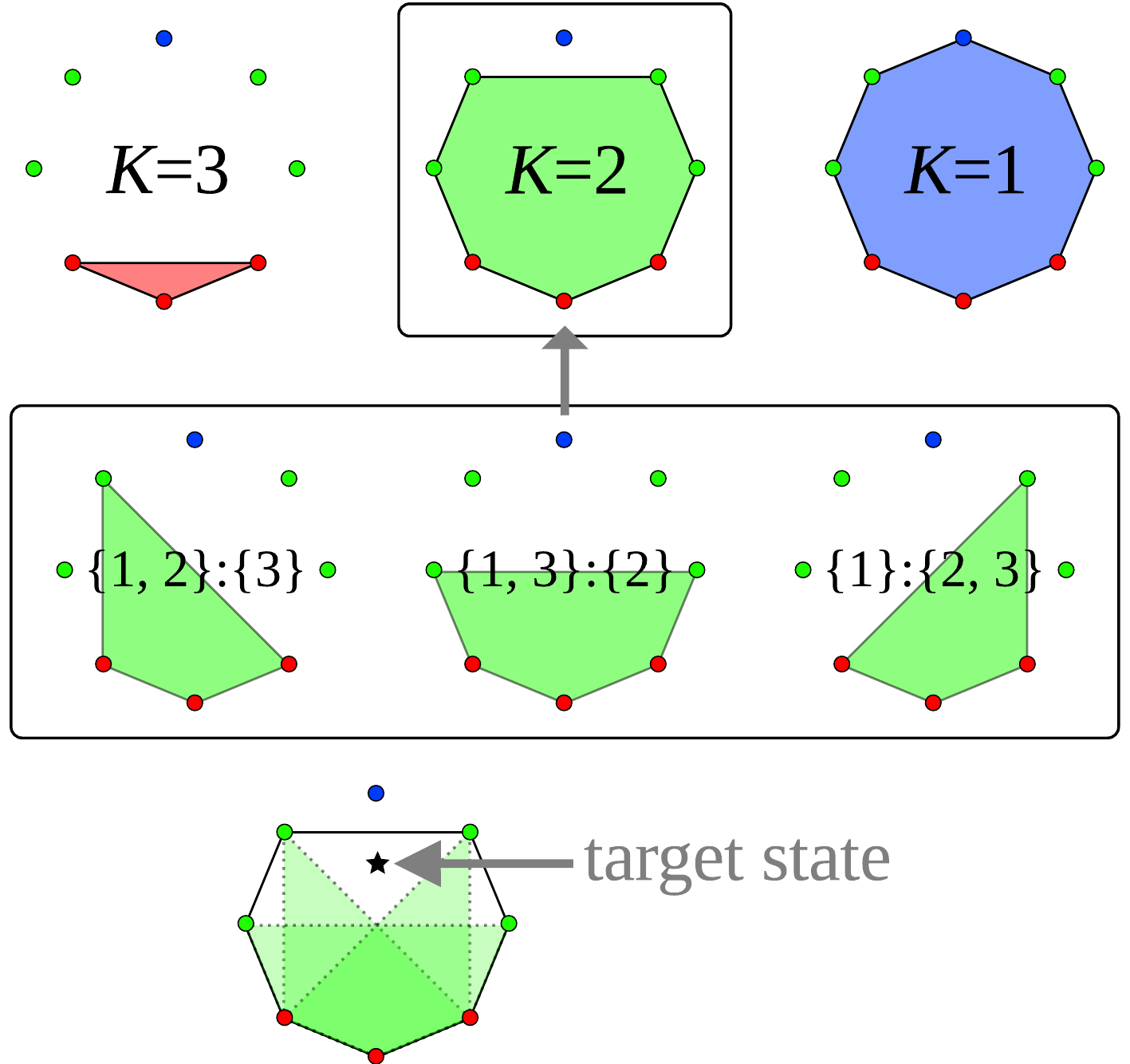}
		\caption{(Color online)
			Partitionings of a tripartite system.
			The circles,~$\boldsymbol\circ$, indicate pure state representatives for different notions of separability/entanglement (distinguished by the lightness[color]).
			The top row depicts the convex sets of $K$-partitions.
			The $K{=}2$ case is a convex combination of the individual bipartitions $\mathcal I_1{:}\mathcal I_2$, which are given in the pattern in the middle row.
			The bottom shows the overlay of the individual bipartitions (dashed bordered sets) as an inset into the convex set of all bipartitions (solid borders).
			The target state~$\boldsymbol{\star}$ is a convex combination of 2-partite separable states. It is therefore not genuinely entangled, but 3-entangled.
		}\label{Fig:Structure}
	\end{figure}

	Figure~\ref{Fig:Structure} shows the different kinds of separability (or entanglement) in the tripartite scenario, $N=3$.
	The circles represent pure states, which are the extremal points of the convex hull of separable states with respect to given partitionings.
	A state lying outside of these sets are entangled in that particular notion.
	For $K=3$, we have the statistical mixture of pure, fully separable states $|s_{\{1\}{:}\{2\}{:}\{3\}}\rangle$ (red, left pattern in top row of Fig.~\ref{Fig:Structure}).
	For $K=1$, we trivially get all states (blue, right pattern).

	The partitions for $K=2$ (green) consist of three individual bipartitions, shown in the middle row of Fig.~\ref{Fig:Structure}.
	The (green) circle on the top-left corresponds to a pure state $|s_{\{1,2\}{:}\{3\}}\rangle$, which is not of the form $|s_{\{1\}{:}\{2\}{:}\{3\}}\rangle$, $|s_{\{1,2\}{:}\{3\}}\rangle$, or $|s_{\{1,3\}{:}\{2\}}\rangle$.
	Thus, it is entangled with respect to those notions of separability.
	Similarly, the (green) circle on the top-right and the two (green) middle circles are exclusively separable with respect to the individual partition $\{1\}{:}\{2,3\}$ and $\{1,3\}{:}\{2\}$, respectively.
	All of them are certainly $K=2$-separable in the convex combination of all bipartitions.

	In the bottom row of Fig.~\ref{Fig:Structure}, we also show the three {\it individual} bipartitions $\{1,2\}{:}\{3\}$, $\{1,3\}{:}\{2\}$, and $\{1\}{:}\{2,3\}$ (dashed borders) included into the {\it convex combination} of biseparable states (solid border).
	Our target state (indicated by a star) is entangled with respect to all individual bipartitions, but it is separable with respect to the convex combination of bipartitions.
	These states are particularly interesting as they are not genuine multipartite entangled.
	We will show explicitly for a six-partite system, $N=6$, that such states still exhibit rich multipartite entanglement properties.

\section{Entanglement criteria for convex combinations}
	A criterion for detecting entanglement of states is based on the  entanglement witnesses~\cite{HHH96}. 
	It can be formulated as follows:
	A state is entangled if a Hermitian operator $\hat{L}$ exists, whose expectation value is smaller than the minimal attainable value for all separable states $\hat\sigma$~\cite{SV13},
	\begin{equation}
		\langle\hat{L}\rangle<\inf_{\hat{\sigma}}\{\text{tr}(\hat{L}\hat{\sigma})\}.
		\label{Eq:Generalcriterion}
	\end{equation}
	This criterion is general and covers any kind of inseparability. 
	It applies therefore to either individual partitions or convex combinations.
	Using this result and the following property,
	\begin{equation}
		\inf\left\{\int dP(x) x\right|\left.\int dP{=}1\text{ and } P{\ge}0\right\}=\inf_{x}\{x\},
		\label{Eq:Convex}
	\end{equation}
	we derive the lower bound in~\eqref{Eq:Generalcriterion} for $K$-separable states:
	\begin{align*}
		&\inf_{\hat{\sigma}_K}\{\text{tr}(\hat{L}\hat{\sigma}_K)\}
		\stackrel{\eqref{Eq:KSeparable}, \eqref{Eq:Convex}}{=}
		\min_{\mathcal{I}_1{:}\ldots{:}\mathcal{I}_K}\inf_{\hat{\sigma}_{\mathcal{I}_1{:}\ldots{:}\mathcal{I}_K}}
		\{\text{tr}(\hat{L}\hat{\sigma}_{\mathcal{I}_1{:}\ldots{:}\mathcal{I}_K})\}
		\\&\stackrel{\eqref{Eq:IndPartSep}, \eqref{Eq:Convex}}{=}
		\min_{\mathcal{I}_1{:}\ldots{:}\mathcal{I}_K}\inf{}_{|s_{\mathcal{I}_1{:}\ldots{:}\mathcal{I}_K}\rangle}
		\{\langle s_{\mathcal{I}_1{:}\ldots{:}\mathcal{I}_K}|\hat{L}|s_{\mathcal{I}_1{:}\ldots{:}\mathcal{I}_K}\rangle\}.
	\end{align*}
	The equation labels over the equal signs indicates that those equations have been used for rewriting.
	
	The minimization of $\langle s_{\mathcal{I}_1{:}\ldots{:}\mathcal{I}_K}|\hat{L}|s_{\mathcal{I}_1{:}\ldots{:}\mathcal{I}_K}\rangle$ for pure, $\mathcal{I}_1{:}\ldots{:}\mathcal{I}_K$-separables states has been treated in Ref.~\cite{SV13}. 
	There, so-called ``separability eigenvalue equations'' have been derived. 
	The solution of those equations for a given observable $\hat L$ yields the minimal separability eigenvalue $g^{\text{min}}_{\mathcal{I}_1{:}\ldots{:}\mathcal{I}_K}$, which is also the desired infimum for separable states of the individual $K$-partition $\mathcal{I}_1{:}\ldots{:}\mathcal{I}_K$.
	We can conclude:
	A state is inseparable with respect to the convex combination of all $K$-partitions ($K$-entangled), if and only if there exists a Hermitian operator $\hat{L}$, such that
	\begin{equation}
		\langle \hat{L}\rangle<g_K^{\min}=\min_{\mathcal{I}_1{:}\ldots{:}\mathcal{I}_K}\{g^{\text{min}}_{\mathcal{I}_1{:}\ldots{:}\mathcal{I}_K}\}.
		\label{Eq:Kentanglementcondition}
	\end{equation}
	Although this condition clearly differs from the approach for individual partitions~\cite{GSVCRTF15}, it is remarkable that we can use a similar calculus.
	The method of separability eigenvalues was introduced to uncover entanglement of individual $K$-partitions $\mathcal{I}_1{:}\ldots{:}\mathcal{I}_K$, via $\langle \hat{L}\rangle<g^{\text{min}}_{\mathcal{I}_1{:}\ldots{:}\mathcal{I}_K}$~\cite{SV13}.
	It serves now for detecting entanglement among convex combinations of all $K$-partitions [inequality~\eqref{Eq:Kentanglementcondition}].

\section{Witnessing multimode Gaussian states}
	A Gaussian state is fully described by its covariance matrix.
	In the following, we will use the notation
	\begin{equation}
		\mathbf{{\hat{\xi}}}=(\hat{x}_1,\ldots,\hat{x}_N,\hat{p}_1,\ldots,\hat{p}_N)^T
		\label{Eq:quadratures}
	\end{equation}
	for a vector containing the amplitude ($\hat x_j$) and phase ($\hat p_j$) quadratures of all possible modes ($j=1,\ldots,N$).
	The covariance matrix $C$ of a Gaussian state can be written in terms of the symmetrically ordered elements   $C^{ij}=\langle\hat{\xi}_i\hat{\xi}_j+\hat{\xi}_j\hat{\xi}_i\rangle/2-\langle\hat{\xi}_i\rangle\langle\hat{\xi}_j\rangle$.
	As local displacements do not affect the entanglement, it is sufficient to analyze the covariance matrix of a Gaussian state, assuming $\langle \hat \xi_j\rangle=0$.
	Thus, the most general form of a Gaussian test operator $\hat L$ is the quadratic combination
	\begin{align}
		\hat{L}&=\sum_{i,j=1}^{2N}M_{ij}\hat\xi_i\hat\xi_j,
		\label{Eq:testoperator}
	\end{align}
	with a symmetric, positive definite $2N\times2N$-matrix $M=(M_{ij})_{i,j=1}^{2N}$.
	Note that Williamson's theorem allows us to diagonalize such a matrix $M$ into a form ${\rm diag}(\lambda_1,\ldots,\lambda_N,\lambda_1,\ldots,\lambda_N)$ in terms of symplectic operations, see, e.g.,~\cite{SCS99}.

	The minimal separability eigenvalue of $\hat L$ in Eq.~\eqref{Eq:testoperator} for an individual partition $\mathcal{I}_1{:}\ldots{:}\mathcal{I}_K$ is given by~\cite{GSVCRTF15}:
	\begin{equation}
		g^{\min}_{\mathcal{I}_1{:}\ldots{:}\mathcal{I}_K}=
		\sum_{j=1}^K\sum_{k=1}^{|\mathcal{I}_j|}\lambda_k^{\mathcal{I}_j},
		\label{Eq:minSEVal}
	\end{equation}
	where $|\mathcal{I}_j|$ is the cardinality of $\mathcal I_j$, and $\lambda_k^{\mathcal{I}_j}$ are the diagonal values of the Williamson decomposition of the sub-matrix which solely consists of the rows and columns of $M$ that are in the index set $\mathcal{I}_j$ (see also the Supplemental Material of Ref.~\cite{GSVCRTF15}). 
	Finally, the entanglement condition~\eqref{Eq:Kentanglementcondition} is given by the bound
	\begin{equation}
		g^{\min}_K=\min_{\mathcal{I}_1{:}\ldots{:}\mathcal{I}_K}\left\{
		g^{\min}_{\mathcal{I}_1{:}\ldots{:}\mathcal{I}_K}\text{ in Eq.~\eqref{Eq:minSEVal}}\right\}.
		\label{Eq:KminSEVal}
	\end{equation}
	Hence, we have formulated an infinite number (for any positive, symmetric matrix $M$) of multipartite $K$-entanglement probes in an analytical form. 
	This includes as a subclass Gaussian tests for genuine entanglement, $g_{K=2}^{\min}>\langle \hat L\rangle$, which have been recently studied~\cite{SL15}.

	In order to get the best entanglement signature of all test operators $\hat L$ in terms of matrices $M$ [Eq.~\eqref{Eq:testoperator}], we take the analytical solutions in Eqs.~\eqref{Eq:minSEVal} and~\eqref{Eq:KminSEVal} and numerically minimize the signed significances:
	\begin{equation}
		\Sigma_{\mathcal{I}_1{:}\ldots{:}\mathcal{I}_K}=\frac{\langle\hat L\rangle-g^{\min}_{\mathcal{I}_1{:}\ldots{:}\mathcal{I}_K}}{\Delta\langle\hat L\rangle}
		\text{ and }
		\Sigma_{K}=\frac{\langle\hat L\rangle-g^{\min}_{K}}{\Delta\langle\hat L\rangle},
		\label{Eq:signedsignificance}
	\end{equation}
	where $\Delta\langle\hat L\rangle$ denotes the experimental error of $\langle \hat L\rangle$, by finding the optimal matrix $M$ for each of those significances.
	The signed significance is negative, $\Sigma_\chi<0$, if the state is entangled with respect to the given notion of separability, $\chi=K$ or $\chi=\mathcal{I}_1{:}\ldots{:}\mathcal{I}_K$, which is certified with a significance of $|\Sigma_\chi|$-standard deviations. 
	The numerical minimization was performed with a genetic algorithm~\cite{GSVCRTF15} which can, in principle, not only find local minima, but also global ones~\cite{HH04}.
	Hence, one could claim that a positive value $\Sigma_\chi$ corresponds to a $\chi$-separable covariance matrix. 
	However, we will more carefully state in such a case that no $\chi$-entanglement can be detected.

\section{Characterization of the SPOPO multimode quantum state}
	The details on the experimental generation and the characterization of the produced state in terms of uncorrelated squeezed ``supermodes'' can be found in Refs.~\cite{ARCFFT14} and~\cite{RMSFT14}.
	The light under study is a femtosecond frequency comb of zero mean value spanning over roughly  $\sim10^{5}$ individual equally spaced frequency components, generated by parametric down conversion of a pump frequency comb in a synchronously pumped optical parametric oscillator (SPOPO).

	It is analyzed through a series of balanced homodyne detections that use different pulse shaped local oscillators of adjustable spectrum.
	This allowed us to experimentally determine the full covariance matrix $C$, containing the noise variances in different frequency bands covering the whole spectrum of the SPOPO state, as well as the correlations between them.
	Assuming a Gaussian distribution of the quantum fluctuations, a reasonable assumption in the present situation, one retrieves in such a way the full information about the generated quantum state, at least within the frequency resolution given by the width of the frequency bands used in the measurements.
	For the experiment described here, the spectrum was partitioned into six bands of equal energy, which allowed us to determine the 144 matrix elements $C^{ij}$ of the $12\times 12$ covariance matrix together with the corresponding experimental uncertainties $\Delta C^{ij}$.

	Uncorrelated supermodes, i.e., well defined combinations of frequency modes corresponding to femtosecond pulses of specific shapes~\cite{PVTF10}, were extracted from the matrix. 
	They have squeezing levels between $-2.6\,{\rm dB}$ and $+3.0\,{\rm dB}$.
	The generated state is clearly mixed, as its purity, ${\rm tr}\hat\rho^2=(\det C)^{-1/2}=86.4\%$, is below one.

\section{Entanglement structure of a six-mode SPOPO state}
	For an $N{=}6$-mode state, 203 possible {\it individual} partitions exist.
	That is one trivial partition $\mathcal I_1=\{1,\ldots,6\}$, 31 bipartitions $\mathcal I_1{:}\mathcal I_2$, 90 tripartitions, 65 four-partitions, 15 five-partitions, and one six-partition $\{1\}{:}\ldots{:}\{6\}$.
	Hence, we have six {\it convex combinations} of $K$-partitions.

	\begin{figure}[ht]
		\centering
		\includegraphics[width=0.45\textwidth]{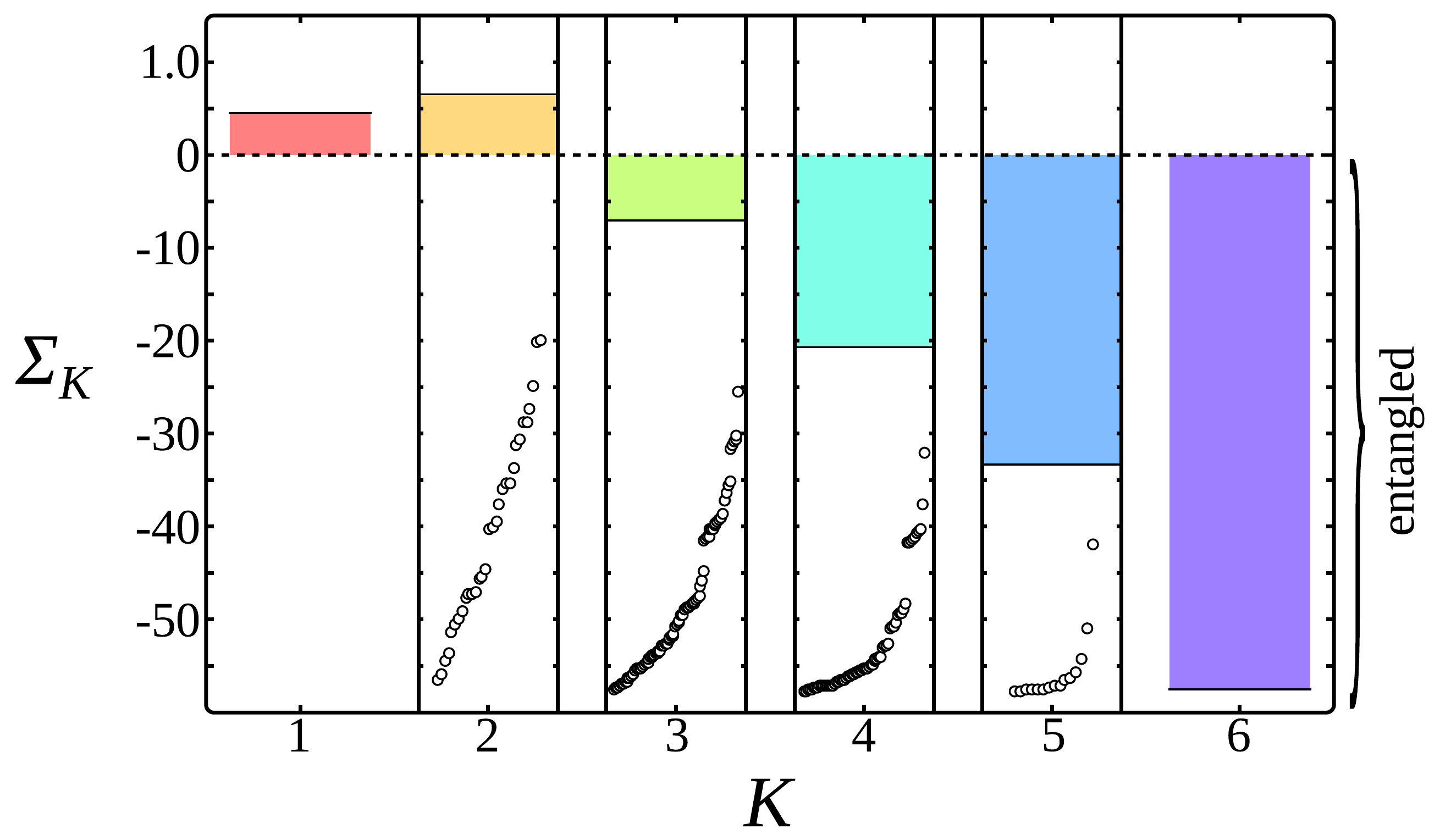}
		\caption{(Color online)
			Signed significance $\Sigma_K$ (bars), for $1\leq K\leq6$, calculated from the data of the SPOPO state.
			The insets for $2\leq K\leq 5$ give the values for the individual partitions, $\Sigma_{\mathcal I_1{:}\ldots{:}\mathcal I_K}$ (circles), sorted in increasing order.
			For better visibility, the positive part of the ordinate has a different scaling than the negative (entangled) part.
			Despite no signature of genuine entanglement, $\Sigma_{2}>0$, the state shows highly significant other forms of multipartite entanglement.
		}\label{fig:Kpartitetest}
	\end{figure}

	The results of our analysis are shown in Fig.~\ref{fig:Kpartitetest} in terms of the minimized signed significances in Eq.~\eqref{Eq:signedsignificance}. 
	The trivial partition $K=1$ yields $\Sigma_{K=1}>0$, which means that the measured covariance is a physical one. 
	The value $\Sigma_{K=2}>0$ shows that \textit{no detectable genuine entanglement exist in the SPOPO quantum frequency comb}.
	Yet, for all $K\geq3$, $K$-entanglement is verified with a significance of at least seven standard deviations, $\Sigma_{K>2}<-7$.
	Such types of multipartite entanglement are not accessible with entanglement probes that are only sensitive to genuine entanglement. 

	Considering the circles in the insets for $1\leq K\leq 6$ in Fig.~\ref{fig:Kpartitetest}, it can be seen that the same six-mode state is entangled with respect to all nontrivial, individual partitions -- even for $K=2$.
	Therefore, the SPOPO state is entangled with respect to any individual bipartition, even though it cannot be identified as a genuinely entangled state:
	The subtle structures of multipartite entanglement are invisible for genuine entanglement probes.

	Here, we see clearly that entanglement of some or even all individual partitions $\mathcal I_1{:}\ldots{:}\mathcal I_K$ of the length $K$ does not necessarily imply $K$-entanglement.
	Rather, it is the convex combination of the individual partitions that is responsible for the separability or inseparability. 
	The inverse, however, is true:
	$K$-entanglement implies entanglement with respect to all individual $K$-partitions.
	This follows from the condition~\eqref{Eq:Kentanglementcondition} by taking a proper test operator $\hat L$ for the convex combination and the same $\hat L$ for every individual $K$-partition, as $g^{\min}_K\leq g^{\min}_{\mathcal{I}_1{:}\ldots{:}\mathcal{I}_K}$. 
	Let us finally stress that this approach can be extended to study other convex combinations of some individual partitions which are not limited by a fixed $K$-value.

\section{Summary and conclusion}
	We have studied different forms of $K$-party entanglement in multimode states. 
	An analytical approach to construct the corresponding entanglement tests was derived and further elaborated for covariance based entanglement probes. 
	To optimize over the resulting infinite set of all analytical Gaussian witnesses, a numerical optimization was performed. 
	This approach allows us to classify entanglement in Gaussian states with an arbitrary number of modes.

	We then focused on the characterization of $K$-partite entanglement of multi-mode Gaussian states, that we applied to a parametrically generated multimode frequency comb. 
	It was shown for a six-mode example that our system shows an interesting form of entanglement, though it did not exhibit genuine multipartite entanglement.
	That is, the SPOPO state turns out to be a biseparable state which is $K$-entangled for any $K{=}3,\ldots,6$.
	Moreover, we detected entanglement with respect to all individual partitions, even all the individual bipartitions.
	Thus, the absence of genuine entanglement does not give much insight into the entanglement structure.
 
	This work proves the great interest to investigate the nature of entanglement beyond genuine entanglement in highly multipartite systems. 
	A lot of questions remain to be investigated concerning other possible types of multipartite entanglement and in particular their relation to quantum tasks in various types of quantum computing protocols. 
	Our construction of more general entanglement criteria, likely to access multipartite quantum correlations beyond bipartitions, provides a good starting tool to tackle such problems.

	As a comment to the Einstein-Podolsky-Rosen paradox~\cite{EPR35}, Schr\"odinger emphasized that a compound quantum system includes more information than provided by the individual subsystems~\cite{SCH35}.
	Considering our scenario at hand, we may extend such a statement.
	Namely, multipartite entanglement is much richer than one can infer from the genuine entanglement of bipartitions.

\paragraph*{Acknowledgement}
	This work has received funding from the European Union's Horizon 2020 research and innovation program under grant agreement No 665148.

\end{document}